\documentclass[useAMS,usenatbib]{mn2e}
\usepackage[dvips]{color}
\usepackage{graphicx,natbib}
\def\ms{\mbox{$M_\odot$}}
\def\rc{\mbox{$R_{\rm c}$}}
\def\rt{\mbox{$R_{\rm t}$}}

\title[Multi-Conjugate Adaptive Optics VLT imaging of the distant 
old open cluster FSR1415]{Multi-Conjugate Adaptive Optics VLT imaging of the distant 
old open cluster FSR1415}
\author[Y. Momany, S. Ortolani, C. Bonatto, E. Bica and B. Barbuy]
{Y. Momany$^{1,2}$\thanks{E-mail: yazan.almomany@oapd.inaf.it},
{S. Ortolani$^{3}$},
{C. Bonatto$^{4}$},
{E. Bica$^{4}$},
{B. Barbuy$^{5}$}\\
$^{1}$INAF: Osservatorio Astronomico di Padova, Vicolo dell'Osservatorio 5, 
	I-35122 Padova, Italy\\
$^{2}$ESO: European Southern Observatory, Alonso de Cordova 3107, Santiago, 
	Chile\\
$^{3}$Universit\`a di Padova, Dipartimento di Astronomia, Vicolo
 	dell'Osservatorio 2, I-35122 Padova, Italy\\
$^{4}$Universidade Federal do Rio Grande do Sul, Departamento de Astronomia
CP 15051, RS, Porto Alegre 91501-970, Brazil\\
$^{5}$Universidade de S\~ao Paulo, IAG, Rua do Mat\~ao 1226,
Cidade Universit\'aria, S\~ao Paulo 05508-900, Brazil
}
\begin{document}

\date{Accepted  2008 September  29.   Received 2008  September 28;  in
original form 2008 August 16}

\pagerange{\pageref{firstpage}--\pageref{lastpage}} \pubyear{2008}

\maketitle

\label{firstpage}

\begin{abstract}
We  employ the    recently  installed  near  infrared  Multi-Conjugate
Adaptive Optics demonstrator (MAD)  to determine basic properties of a
newly identified, old and distant, Galactic open cluster (FSR\,1415).
The MAD facility remarkably approaches the diffraction limit, reaching
a resolution of $0\farcs07$ (in $K$), that is  also uniform in a field
of $\sim1\farcm8$ in diameter.
The MAD   facility  provides  photometry that  is  $50\%$  complete at
$K\sim19$.  This corresponds to about $2.5$ mag below the cluster main
sequence turn-off.
This  high  quality   data  set  allows  us  to   derive  an  accurate
heliocentric distance of $8.6$ kpc,  a metallicity close to solar, and
an age of $\sim2.5$\,Gyr.  On the other hand, the deepness of the data
allow  us  to   reconstruct  (completeness-corrected)  mass  functions
indicating a relatively massive cluster, with a flat core MF.
The VLT/MAD  capabilities will therefore   provide fundamental data in
identifying/analyzing  other  faint and  distant open  clusters in the
Galaxy III and IV quadrants.
\end{abstract}

\begin{keywords}
instrumentation: adaptive optics --
infrared: stars --
techniques: photometric --
Stars: Population II --
Galaxy: open clusters and associations --
Individual: FSR\,1415
\end{keywords}

\section{Introduction}

Open  clusters are continuously  affected  by dynamical processes such
as: {\em (i)}  mass loss, linked to stellar  evolution, mass segregation and
evaporation; {\em (ii)} tidal interactions with the Galactic disk and bulge;
and  {\em (iii)} collisions  with giant  molecular  clouds.  These processes
combined tend to accelerate the internal dynamical evolution which, as
the  cluster  ages, results in  significant   changes  in the  cluster
structure   and  mass  function  (\citealt{OldOCs},    and  references
therein).  
Eventually, the majority of the open clusters will be dispersed in the
Galactic    stellar  field,  or   become  poorly-populated    remnants
(\citealt{PB07} and references therein).  In fact, open clusters older
than  $\sim1$\,Gyr are found preferentially  near the Solar circle and
in  the   outer Galaxy (e.g.   \citealt{Friel95}; \citealt{DiskProp}),
where the frequency  of dynamical  interactions  with giant  molecular
clouds and the disk is low (e.g.  \citealt{Salaris04};
\citealt{Upgren72}).

As a reflex  of the above scenario, less  than $\sim18\%$ of the  open
clusters listed   in WEBDA are  older  than 1\,Gyr, with  most of them
located outside the Solar circle (see, e.g., Fig.~1 of
\citealt{OldOCs}).  Granted the above   aspects,  the discovery and
derivation  of astrophysical   parameters of   old open  clusters   is
expected to   result  in  a better    understanding  of the  dynamical
processes  that affect star clusters,  with reflexes on star formation
and evolution processes, dynamics of  N-body systems, and the geometry
of the Galaxy, among others.

In this framework,  the construction of  an old ($\ge1$  Gyr), distant
and complete open cluster sample acquires particular importance.
An effort  in this  direction   is the new   catalogue of  $1021$ star
cluster candidates that was   recently presented by Froebrich  et  al.
(2007a - hereafter FSR objects). It spans all Galactic longitudes (for
$|b|<20^\circ$), and  is   based on an  over-density  automated search
algorithm applied to the 2MASS\footnote{The Two Micron All Sky Survey,
available                                   at\\                    {\em
www.\-ipac.\-caltech.\-edu/\-2mass/\-releases/\-allsky/ }}   database.
Examination of this  catalogue has surprisingly allowed the  detection
of   new  globular  clusters  (FSR\,1735, \citealt{FMS07};  FSR\,1767,
\citealt{FSR1767}) and a number of old open clusters (e.g.  FSR\,1744,
FSR\,89, FSR\,31, \citealt{BB07}).
We have examined  many  cluster  candidates from this  catalogue,  and
verified that several of these must have  the turn-off (TO) much below
the 2MASS limit.   Thus, and for a  few cases, the photometry provided
by a 4m-class telescope was still not deep enough (\citealt{Fmd08};
\citealt{FMS07}), and consequently, 8m-class telescopes (provided that
adaptive optics is available) were needed.

\begin{table}
\centering
\caption[]{Log of the MAD observations obtained on January
13$^{th}$ 2008.}
\label{tab1}
\renewcommand{\tabcolsep}{2.mm}
\renewcommand{\arraystretch}{1.5}
\begin{tabular}{lllll}
\hline
Filter & FWHM & airmass & DIT (sec.) & NDIT\\
\hline
$K$	& $0\farcs085$	    &  1.145 & 10 & 27\\
$K$	& $0\farcs073$	    &  1.137 & 10 & 27\\
$K$	& $0\farcs073$	    &  1.129 & 10 & 27\\
$K$	& $0\farcs074$	    &  1.122 & 10 & 27\\
\hline			     	        
$J$	& $0\farcs145$ 	    &  1.248 & 10 & 27\\
$J$	& $0\farcs101$      &  1.234 & 10 & 27\\
$J$	& $0\farcs176$      &  1.221 & 10 & 27\\
$J$	& $0\farcs195$      &  1.209 & 10 & 27\\
\hline
\end{tabular}
\end{table}

The     recent  availability  of  a    Multi-Conjugate Adaptive Optics
Demonstrator (MAD)  at the VLT UT3-Melipal,  and  the public  call for
Science Verification encouraged us to propose $JK$ observations of the
faint cluster FSR\,1415.
Indeed, the inspection  of the 2MASS colour-magnitude diagrams  (CMDs)
indicated the presence of either  (i) a rich red clump;  or (ii) a red
horizontal branch.  Disentangling between these two, similar, features
allows one to infer the  detection of a distant  old open cluster or a
globular  cluster, respectively.  The  results presented in this paper
(Sect.  2)  unequivocally point towards the  detection  of an old open
cluster.
We stress that there are very few photometric studies of Galactic star
clusters with standard Adaptive Optics (Trumpler~14,
\citealt{Ascenso2007})   or  with  Multi-conjugate    Adaptive  Optics
(Trapezium by  \citealt{Bouy08}, see also  the case of the UKS2323-326
dwarf galaxy by \citealt{gull08}).
FSR\,1415  is a low latitude  target  projected almost orthogonally to
the     Galactic     anti-centre    direction  ($\ell=263.74^{\circ}$,
$b=-1.81^{\circ}$).   The     J2000   equatorial    coordinates   are:
($\alpha,\delta$)$=$($8^{\rm           h}:40^{\rm           m}:24^{\rm
s}$,$-44^{\circ}:43^{\rm m}:05^{\rm s}$).

Although  the present study addresses a  single open  cluster, it lays
the basis for  the identification  of  a complete sample of  old  open
clusters in the IV Galaxy quadrant, known for  a dearth of distant and
old clusters.
Ultimately, a complete  sample will  be  useful for understanding  the
evolution of the disk, in  particular that concerning radial gradients
and age-metallicity relations (\citealt{Friel95}; \citealt{Rocha06};
\citealt{Maciel07}).
In  Sect.~2 we describe  the instrument and  observing procedures.  In
Sect.~3  we present the    photometry and derive   cluster parameters,
including a study of the radial structure.  In Sect.  4 we discuss the
results of FSR\,1415 in the context of other  old open clusters and the
properties of the disk. In Sect. 5 conclusions and perspective work on
clusters with adaptive optics are presented.

\begin{figure}
\centering
\includegraphics[width=9cm]{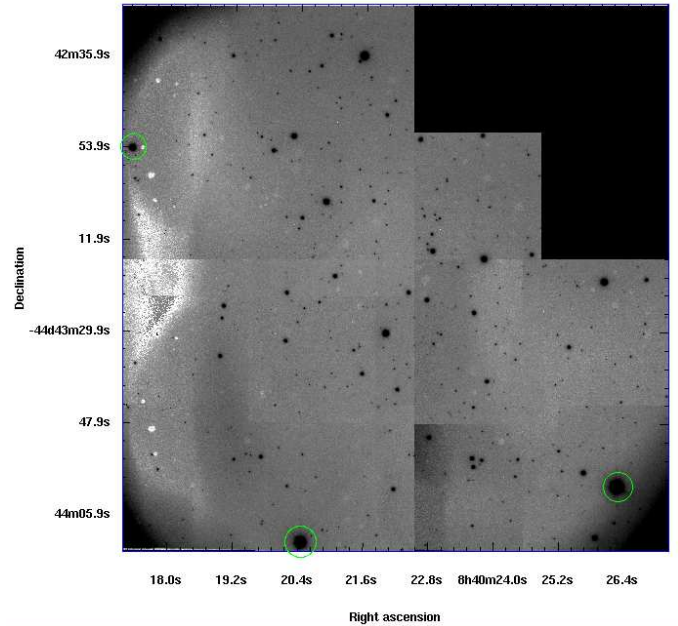}
\caption{The MAD mosaic of the 8 $J$ and $K$ dithered images of
FSR\,1415, East  to the right  North  to the top. The  overall  circular
field  of view is $\sim1.8\arcmin\times1.8\arcmin$.  The irregular shape of
the mosaic is  due to the  lacking of a  fifth dithered image.  Border
vignetting and other features (due  to the movable scanning table) are
unavoidable.}
\label{f_fov}
\end{figure}
\begin{figure}
\centering
\includegraphics[width=9.cm]{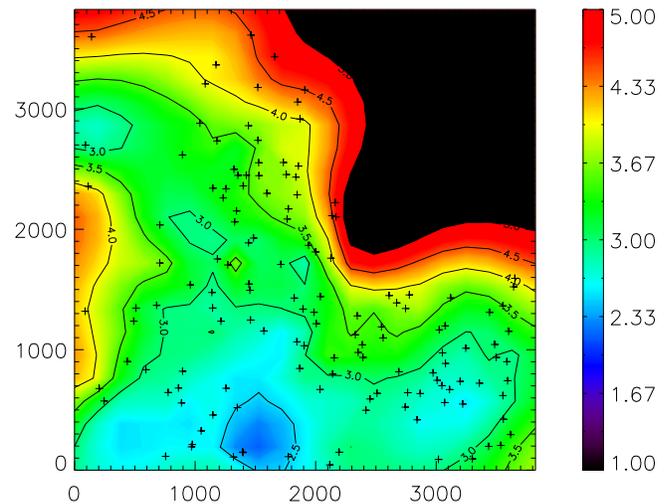}
\caption{The X and Y FWHM map  (in pixels) of the  8 $J$ and $K$
mosaiced images. East to the right, North to the top. Crosses show the
stars  identified and   used in  the  FWHM  computation  (1  pixel  is
$0\farcs028$).}
\label{f_FWHM}
\end{figure}

\section{Observations}
\label{Observ}

MAD was developed by the ESO Adaptive Optics  Department to be used as
a visitor instrument at Melipal  in view of  an application to ELT. It
was installed at the visitor  Nasmyth focus and introduces the concept
of multiple    reference stars for     layer oriented adaptive  optics
corrections. 
This allows  a much  wider and more  uniform corrected field  of view,
providing  larger  average  Strehl  ratios  and making  the  system  a
powerful diffraction limit imager.   This is particularly important in
crowded  fields where the  photometric accuracy  is needed.   For more
details  see   {\sc  http://www.eso.org/projects/aot/mad/},   and
\citet{march07}.   
Indeed, the on-sky MAD commissioning of an $\omega$Cen field showed an
enormous gain in angular  resolution and light concentration such that
very  faint  stars  ($K\sim20.5$)  were  easily  distinguished  (at  a
$3\sigma$ level) in just $600$~sec. exposures.
In  October 2007 we  answered a call  for science
demonstration  of   MAD,  and  among  other   applications  (see  {\sc
http://www.eso.org/sci/activities/vltsv/mad/}) we were granted a total
of 2 hours aimed at deriving  the main properties of a distant cluster
candidate (FSR\,1415).

\begin{figure}
\centering
\includegraphics[width=8.cm,height=4cm]{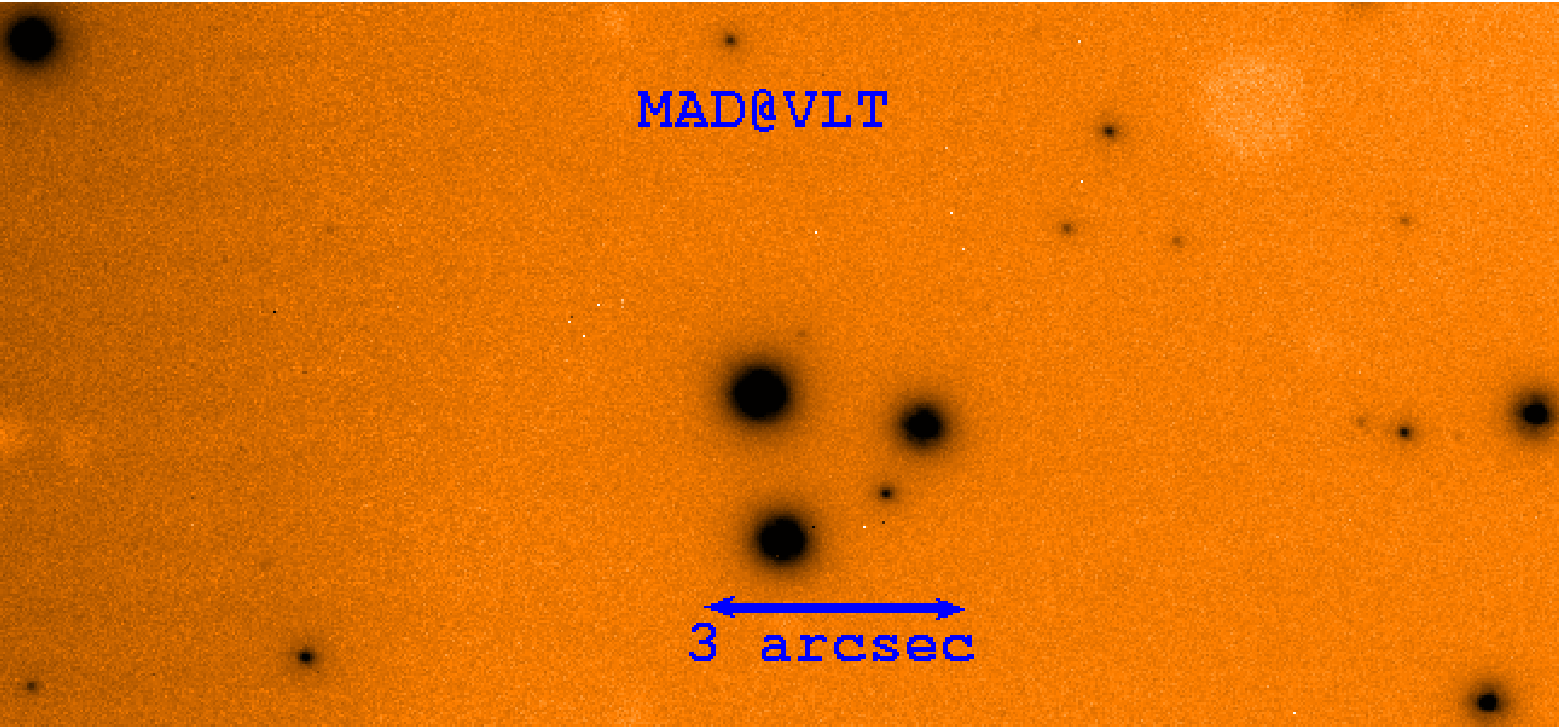}
\includegraphics[width=8.cm,height=4cm]{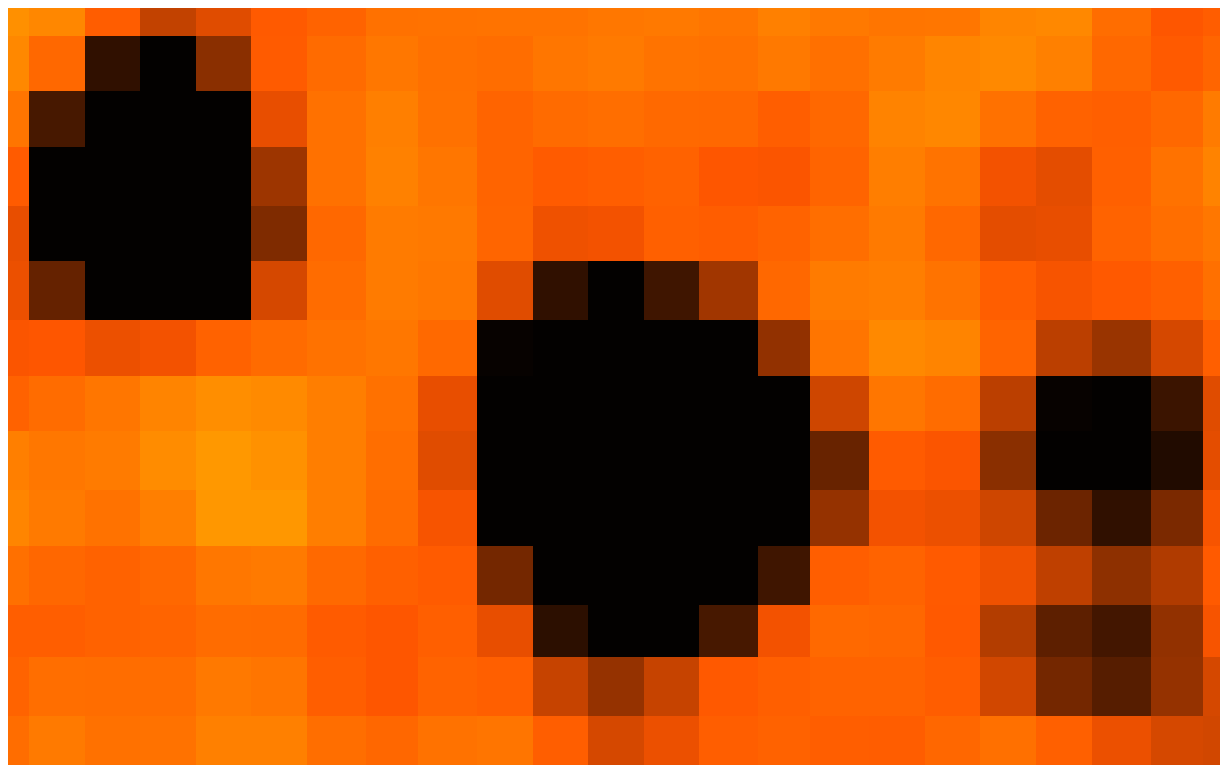}
\caption{The resolution supremacy of the MAD $K$ image
as compared  to 2MASS. The quadruple MAD  stars with $K$ magnitudes of
$13.536$, $13.868$, $14.391$ and $17.155$ are  identified as one single
star ($K=12.853\pm0.040$ and $J=13.536\pm0.038$) in the 2MASS catalogue.
%
}
\label{f_compare}
\end{figure}
%

Within the field  of view of  the  instrument, three reference  bright
stars were selected to ensure the optics correction.
These stars had $V$ magnitudes that ranged between $12.24$ and $13.6$,
according  to their GSC2.2   magnitude system.  While the site  seeing
during the observations was $\approx0\farcs8$,  a diffraction limit in
$K$ of $0\farcs07$ was reached.
Table~\ref{tab1} displays the logbook of the $J$  and $K_{\rm s}$ (for
brevity we use $K$) observations.  The MAD infrared scientific imaging
camera  is based  on    a $2048\times~2048$ pixel    HAWAII-2 infrared
detector with a pixel scale of $0\farcs028$.
Given that the  Detector Integration Time (DIT) is  the amount of time
during which the signal is  integrated onto the detector diodes, while
NDIT  is the  number of  detector integrations  that are  obtained and
averaged together, it follows that the total exposure time of a single
raw  image is  simply  NDIT$\times$DIT.  In our  case,  we obtained  4
dithered images  in each  filter, where DIT  was fixed to  10~sec. and
NDIT to 27. Thus, the total  exposure time is $1080$ sec. (18 min.) in
$J$ and $K$.
The images were
dark and  sky-subtracted and then  flat-fielded following the standard
near infrared  recipes (see for  example \citealt{dutra03}
and \citealt{momany03}), all performed under the {\sc iraf}
environment.

Figure~\ref{f_fov} shows the  mosaic of all  8 $J$  and $K$  images as
constructed by {\sc daophot/montage2} task.  Figure~\ref{f_FWHM} shows
the FWHM  map with an overall  regular distribution across  the entire
field of view with  values ranging between $\sim2.5$ ($0\farcs07$) and
$\sim4.5$ ($0\farcs13$),  where the upper limit is   mostly due to the
fitting process in  border zones with  scarce  stellar population. The
superb VLT/MAD resolution is illustrated in Fig.~\ref{f_compare}, in a
comparison with the 2MASS photometry.

\subsection{Photometric Reduction}

Stellar photometry was obtained by point spread function (PSF) fitting
using  the     tested  {\sc daophot~ii/allframe} (\citealt{stetson94})
package.   ALLFRAME   combines  PSF photometry   carried    out on the
individual  images   and allows the   creation  of  a  master list  by
combining images from different filters, thereby pushing the detection
limit to  fainter  magnitudes.   Once the   FIND and  PHOT  tasks were
performed, we searched  for isolated stars to build  the PSF for  each
single image.  The final PSF was generated with  a PENNY function that
had a quadratic dependence on position in the frame.

\subsection{Calibration}
\begin{figure}
\centering
\includegraphics[width=9.cm,height=9cm]{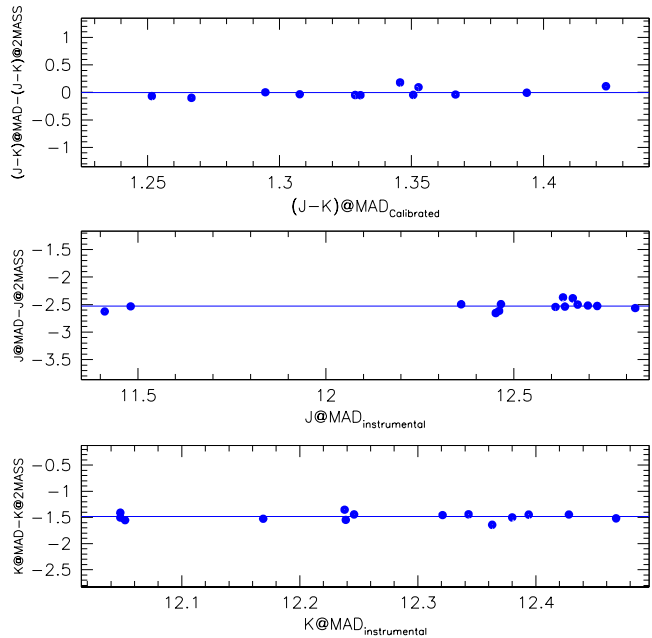}
\caption{The lower and middle panel display a comparison of the $J$
and    $K$   MAD instrumental  photometry  with 2MASS
magnitude.  Upper panel shows the  ($J-K$) colour  difference among the
two photometric systems once the offsets  have been applied to the MAD
data.}
\label{f_2mass}
\end{figure}

Calibration of the $J$  and $K$ data has  been made possible by direct
comparison of the brightest  MAD non-saturated stars with their  2MASS
photometry.
Figure~\ref{f_2mass} shows the  $J$   and $K$  magnitude   differences
between our MAD  instrumental photometry   and  that from  2MASS.   In
particular,  the majority  of the stars  in  common are the red  clump
stars.  From these  stars we estimate  a mean offset of $\Delta~J_{\rm
J@MAD-J@2MASS}=-2.516\pm0.081$         and              $\Delta~K_{\rm
K@MAD-K@2MASS}=-1.477\pm0.074$.  The  upper  panel shows  a comparison
between the MAD and 2MASS photometry  once the above offsets have been
applied to the MAD data.

\subsection{Photometric errors and completeness}
\label{PhotCompl}

\begin{figure}
\centering
\includegraphics[width=9cm,height=9cm]{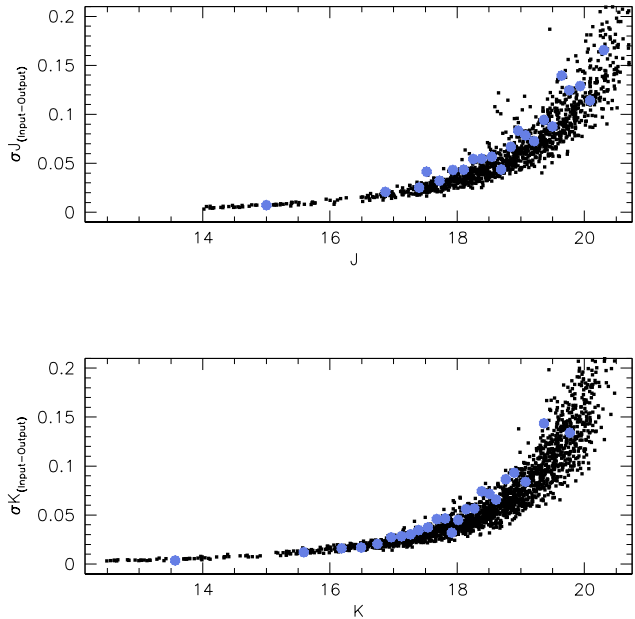}
\caption{The  $J$ and $K$ photometric errors (black dots) as provided
by the DAOPHOT/ALLFRAME  reduction.  Light starred symbols display the
estimated  photometric errors as   derived  from the artificial  stars
experiment.}
\label{f_errors}
\end{figure}

\begin{figure}
\centering
\includegraphics[width=8.cm,height=8cm]{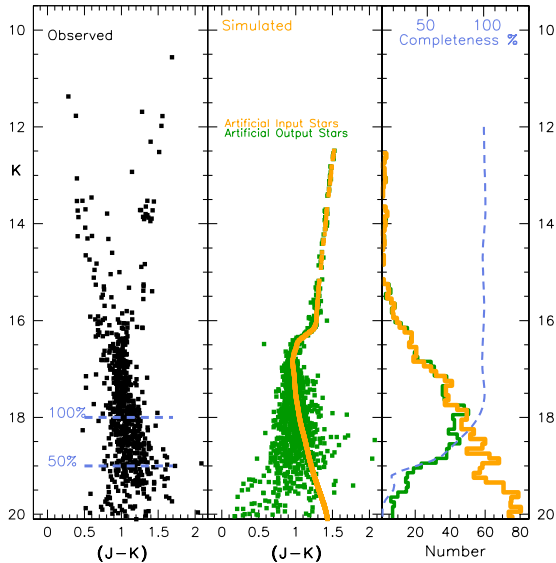}
\caption{Results of the artificial stars experiments. Left panel shows
the observed CMD while  the middle panel  displays
the  input and output CMD  of the simulated stars.
The right  panel compares the luminosity  function of the input/output
artificial stars whereas the light dashed line  shows the computed $K$
completeness level.}
\label{f_completeness}
\end{figure}

Photometric errors and completeness were estimated from the artificial
stars experiments following  procedures similar to  those in \citet{momany02}
and \citet{momany05}.
More than 2300 stars  were added to  the single images, and these were
placed      in  a suitable pattern   of    triangular  meshes to avoid
self-crowding   and   not  closer than   40  pixels   ($1\farcs1$). In
particular, the distribution  of the  $J$  and $K$ magnitudes was  not
casual,  and these were  simulated in  a way to  reproduce the cluster
main sequence and red giant branch loci.
The images with  the artificial stars  added were then re-processed in
the  same manner as the original  images.  The results for photometric
completeness  and errors  are  presented  in  Fig.~\ref{f_errors}  and
Fig.~\ref{f_completeness}.

\begin{table}
\centering
\caption[]{$K$-completeness levels as derived from the artificial
stars experiments for an average ($J-K$)$\sim1.0$ of the input stars.}
\label{t_complet}
\renewcommand{\tabcolsep}{2.mm}
\renewcommand{\arraystretch}{1.5}
\begin{tabular}{ll|ll}
\hline
$K$-mag. & Comp.$(\%)$ & $K$-mag. & Comp. $(\%)$ \\
\hline
   12.0 &  100	 &   18.2  &   89 \\
   12.5 &  100	 &   18.4  &   80 \\
   13.0 &  100	 &   18.6  &   72 \\
   13.5 &  100	 &   18.8  &   62 \\
   14.0 &  100	 &   19.0  &   40 \\
   14.5 &  100	 &   19.2  &   18 \\
   15.0 &  100	 &   19.4  &   20 \\
   15.5 &  100	 &   19.6  &   16 \\
   16.0 &  100	 &   19.8  &   11 \\
   16.5 &   99	 &   20.0  &   10 \\
   17.0 &   98	 &   20.2  &    6 \\
   17.5 &   98	 &   20.4  &    0 \\
   18.0 &   95	 &   ---   &   --- \\
\hline
\end{tabular}
\end{table}

Figure~\ref{f_errors}  compares    the    nominal    {\sc  daophot~ii}
photometric errors (black dots) with those derived from the
comparison  between the magnitude of   input  simulated stars and  the
retrieved magnitudes.   The light   starred symbols  report  the  mean
photometric errors where the  binning is that of a  fixed number of 50
artificial stars.
The  comparison shows an overall  consistency  between the two derived
errors, although there is   a hint of   a slightly higher  photometric
errors as derived from the artificial  stars experiments. On the other
hand,   Fig.~\ref{f_completeness} summarizes the estimated photometric
completeness of our data.  
The left  panel   of  Fig.~\ref{f_completeness} shows   the   observed
colour-magnitude diagram (CMD), whereas the  middle panel displays the
colour-magnitude distribution of the input $\sim2300$ artificial stars
(orange dots obtained reproducing the MS  and RGB fiducial line of the
cluster) and the  retrieved artificial stars (green dots).
The  distribution  of  the  retrieved  artificial  stars reflects  the
increasing  level of  photometric  errors and  the lower  completeness
levels towards fainter magnitudes.
The right panel compares the luminosity  distribution of the input and
output stars, whereas the light dashed  line computes the output/input
stars ratio (reflecting the completeness of the data).  In calculating
this ratio, only stars  recovered in both  the $J$ and $K$  images are
considered.  Thus, Fig.~\ref{f_completeness} shows   that our CMDs are
basically complete down to $K\sim18.0$, reaching a $50\%$ completeness
level around $K\sim19.0$.  For a  mean ($J-K$) colour of $\sim1.0$, we
report the $K$ completeness levels in Table~\ref{t_complet} .

\section{Analysis of the photometry}

\begin{figure}
\centering
\includegraphics[width=8.cm,height=8.cm]{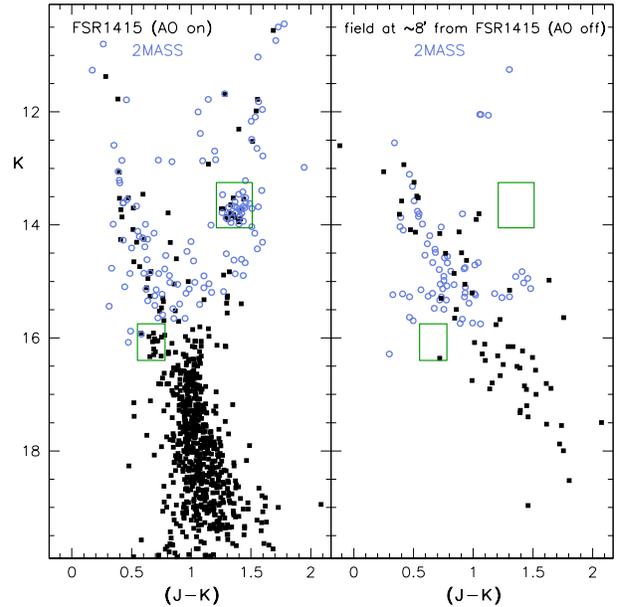}
\caption{Left panel displays the MAD colour-magnitude diagrams of the 
FSR\,1415 field  (adaptive optics   switched on) superimposed  on  the
2MASS photometry (in the same area). Right panel displays the diagrams
of  the SKY field  (adaptive optics  switched  off) at $8\farcm0$ from
FSR\,1415.  The  boxes in both panels highlight   the difference in RC
and BS stars. }
\label{f_cmd}
\end{figure}

The   left     panel of   Figure~\ref{f_cmd}    displays  our  VLT/MAD
colour-magnitude      diagram  of   the     entire  FSR\,1415    field
($\sim1\farcm8\times1\farcm8$).  Superimposed in light open circles 
are 2MASS stars falling in the same field.
The CMD  contains obvious open  cluster features, i.e.   the red giant
branch  (RGB), the red clump  (RC), the main sequence turn-off (MSTO),
and the expected cluster blue stragglers (BS) population.
The  later population is confused  by the  presence of nearby Galactic
disk MS stars (the oblique sequence extending to $K\sim11.0$).
It remains however that the   two prominent open cluster features  are
the RC and the MSTO. Indeed, despite the relatively small MAD field of
view, we sample to about 10 RC stars.
The average location of  these red clump  stars is at $K=13.82\pm0.03$
and ($J-K$)$=1.34\pm0.03$.  On the other hand,  the MSTO is located at
$K=16.65\pm0.05$.  Thus,  the difference between  the  MSTO and RC  is
$\Delta   K^{\rm  TO}_{\rm   RC}=2.83\pm0.06$,  typical   of old  open
clusters.

Although  no  control field has  been  observed, the SKY single images
(usually   medianed  to eliminate  stars)  can be    treated just as a
scientific  image, and reduced  similarly. This should provide us with
an  idea of  the  contaminant  field  population around  FSR\,1415. We
stress the fact however that these SKY images are  not obtained with a
closed adaptive optics (AO)  loop, thus the  resultant CMD  is clearly
shallower.
At  $\sim8\prime$ from the FSR\,1415 one  easily identifies a complete
absence of RC stars. This is also confirmed by the 2MASS photometry in
the same  area, and therefore  the  identification of a  genuine  open
cluster (left panel) is further confirmed.

\subsection{A Blue Stragglers population?}

\begin{figure}
\centering
\includegraphics[width=8.cm,height=8cm]{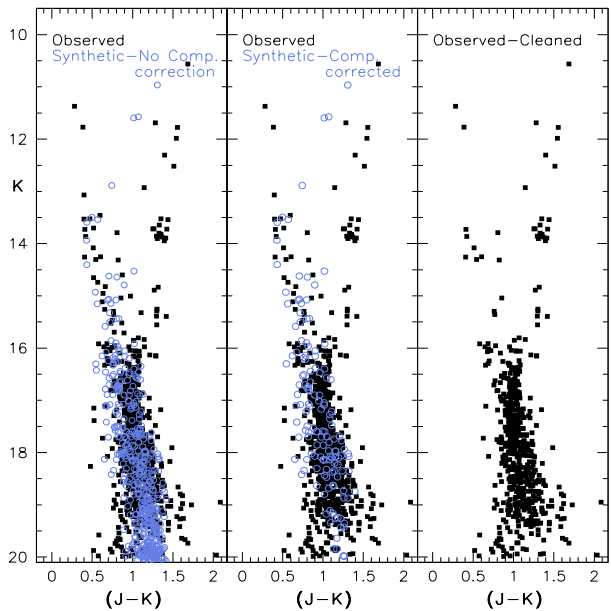}
\caption{Left panel displays the observed MAD CMD along with a
synthetic CMD in  the FSR\,1415 line of  sight. The  middle panel shows
the same diagrams after degrading the synthetic CMD with the estimated
photometric completeness of our   MAD field. Right panel displays  the
``cleaned''  FSR\,1415   diagram   after    subtracting  the      field
contribution.}
\label{f_besanson}
\end{figure}

To confirm the possible presence of a blue stragglers  population
in FSR\,1415 one  needs to account  for the disk population around the
cluster. In the  absence of  a deep  CMD of the  FSR\,1415 surrounding
environment, we attempt  to solve this issue by  the use of  synthetic
diagrams based on models of the Galaxy.
The Besan\c{c}on  (2003) online  simulator provides synthetic  CMDs in
specific  line of  sights  for  a given  area.   Applying the  cluster
Galactic coordinates  and field  of view, in  Fig.~\ref{f_besanson} we
over-plot the synthetic Galactic CMD upon the observed FSR\,1415 CMD.
%

A glance at the left panel of Fig.~\ref{f_besanson} shows an excellent
agreement of  the simulated Galactic population in  this line of sight
and the  observed CMD.  Indeed,  at latitudes of $b=-1.81^{\circ}$ one
expects to detect only young  stellar disk populations and this agrees
with the oblique young MS extending to  $K\sim12.0$.  
There is however a significant  difference  in the star-counts of  the
synthetic MS and that expected in our observed CMD. This is due to the
fact   that the   synthetic    CMDs  do  not ``suffer''    photometric
completeness effects.  Employing  our artificial stars experiments and
the resultant  photometric completeness curve, in  the middle panel we
``correct'' the star-counts    of  the synthetic  MS for   photometric
completeness.  The  middle  panel  shows that  a  remnant  Galactic MS
population  remains present,   and  must be    accounted for  in   any
star-counts analysis of FSR\,1415 populations.
Dividing the CMD in  cells of $0.5$ and  $0.4$ in mag. $K$ and ($J-K$)
respectively, we derive the number of Milky  Way MS stars and randomly
subtract this contribution from the observed MAD diagram.
%

The  right panel of Fig.~\ref{f_besanson} presents  the result of this
statistical decontamination process. Clearly, there remains a clump of
MS stars at $K\sim14.0$ and ($J-K$)$\sim0.5$ and this can be explained
as   due    to  the  small   number   statistics     involved in   the
simulation.  Nevertheless,  
the ``cleaned''  diagram  now  shows a   clear bifurcation  around the
cluster MSTO, indicating the probable detection of the BS sequence.

To  ascertain whether  the  FSR\,1415 BS  candidates are true  cluster
members  we estimate  the  parameter $F_{\rm  RC}^{\rm  BSS}={\rm\log}
~(N_{\rm BS}/N_{\rm RC})$ (see \citealt{demarchi06}) and check if this
frequency is compatible with that in other open clusters,
We  count 14 BS stars   {\it vs} 10 RC  stars,  providing a $F_{\rm RC}^{\rm
BSS}$ of $0.146$. This is fully compatible with  values for other open
clusters ($0.2\pm 0.2$),  indicating that we  have probably detected a
genuine BS population  in FSR\,1415. 
Interestingly, Figure~2  of \citet{momany07} compares
the $F_{\rm RC}^{\rm   BSS}$ in different environment   (spanning from
open and globular  clusters to nearby  dwarf galaxies)  and shows that
open  clusters  and the  lowest  luminosity  dwarf galaxies  (e.g. the
Bootes dwarf at M$_V=-5.5$) share  the  same BS frequency.  This  has
been  interpreted     as  the consequence   of    the  presence of  an
observational upper limit to the frequency production of primordial BS
in low-density stellar systems.

\subsection{Age, metallicity and distance}
\label{age}

\begin{figure}
\centering
\includegraphics[width=8.cm]{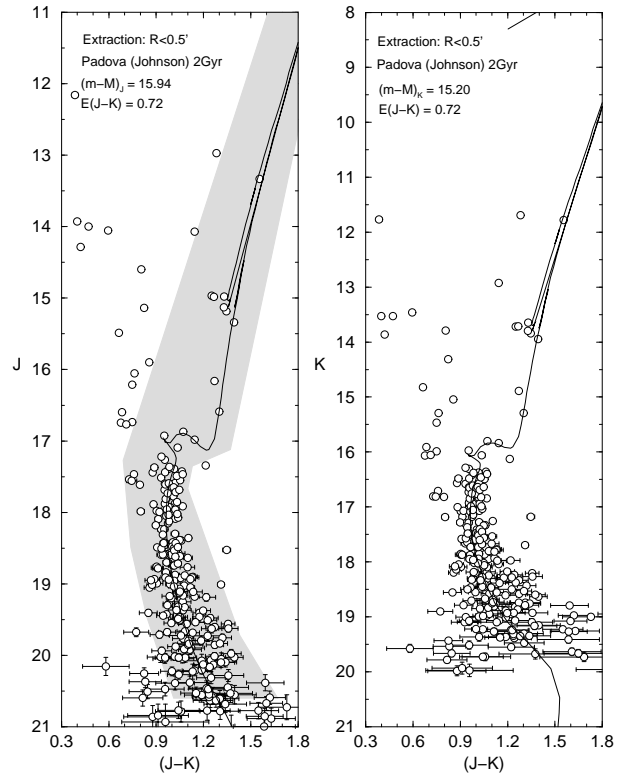}
\caption{Observed $J$ and $K$ {\it vs.} ($J-K$) CMDs of FSR\,1415
within $r<0\farcm5$ from the cluster centre.  Superimposed is the best
fit of   a theoretical $2$ Gyr  solar  metallicity  isochrone from the
Padova  library.   The shaded  area  highlights  the  colour-magnitude
filter used in the derivation of the mass function.}
\label{f_cmds}
\end{figure}

To enhance the signature of the cluster population with respect to the
foreground  disk contamination,  in Fig.~\ref{f_cmds}  we  display the
CMDs of FSR\,1415 extracted within a radius of $r\le0\farcm5$. Despite
the  smaller   field  of  view   the  extracted  CMDs  still   show  a
well-populated RC and  MSTO, ideal features for the  derivation of the
cluster  age.  As  a first  approach, we  assume that  the 4  stars at
$J\approx17$ are cluster sub-giants.  We note indeed, that these stars
survive  our   decontamination  process   (see  the  right   panel  of
Fig~\ref{f_besanson}).    This    granted,   and   following   various
tentatives, in  Fig.~\ref{f_cmds} we show our best  isochrone match to
the  CMD. This  has been  achieved  using isochrones  from the  Padova
library  (\citealt{Girardi02})  for an  age  of  $2$ Gyr,  metallicity
$Z=0.019$,  a  reddening  of  E${\rm  (J-K)}=0.72$,  and  an  apparent
distance modulus ($m-M$)$_{\rm K}=15.20$.
With  respect to the  observed FSR\,1415  RGB and  clump we  note that
isochrones more metal rich than solar are, in general, redder.  On the
other hand,  isochrones with metallicities  less than solar  provide a
poorer match of the cluster MSTO.
Fixing a solar metallicity and allowing for age variations we conclude
that the cluster age is constrained at $2.5\pm0.7$ Gyr.
For infrared ratios between absorption  and colours we use the results
from \citet{DSB2002}, and references therein.
The derived E${\rm (J-K)}$  value thus converts to E${\rm (B-V)}=1.47$
or  A$_{\rm V}=4.56$.    From  the isochrone  fit we  derive  that the
heliocentric  distance    of    FSR\,1415   is   d$_{\odot}=8.59$ kpc.

It would be of great interest to have a spectroscopic determination of
metallicity of FSR\,1415, such  as has been   done in recent years  by
e.g.   Friel  et  al.   (2002), Yong  et   al.  (2005),  Bragaglia  et
al. (2008). This will ultimately reduce the current age uncertainty.
Table~\ref{t_param}  summarizes  the  derived  cluster parameters  and
errors.

\subsection{Location in the Galaxy}

In the derivation of the spatial Galactic coordinates of FSR\,1415 one
needs  to   assume  the  Sun's  distance  from   the  Galactic  centre
(R$_{\odot}$).
Interestingly, the last few years have witnessed a clear trend towards
shorter distance    scales   with   respect  to the nominal     value  of
R$_{\odot}=8.0$ kpc (see \citealt{Reid93}).
Indeed, \citet{Eisenhauer05} derive  R$_{\odot}=7.6\pm0.3$ kpc,
whereas \citet{Nishiyama06} find R$_{\odot}=7.5\pm0.35$  kpc, and \citet{GCProp}
 report  R$_{\odot}=7.2\pm0.3$ kpc.  The   latter value is
based  on an updated   globular   cluster  compilation and   is   adopted
throughout this paper.
We  note   however that recently \citet{Groenewegen08}, with
Population~II  Cepheids and RR Lyrae  stars,  report a longer distance
(R$_{\odot}=7.94\pm0.37$ kpc).
In conclusion, and assuming  R$_{\odot}=7.2\pm0.3$ kpc, we derive that
the Galactocentric distance of FSR\,1415 is $\rm R_{GC}=11.81\pm0.33$\,kpc,
which puts it $\approx4.6$\,kpc outside the Solar circle.

\begin{figure}
\centering
\includegraphics[width=9.cm,height=9cm]{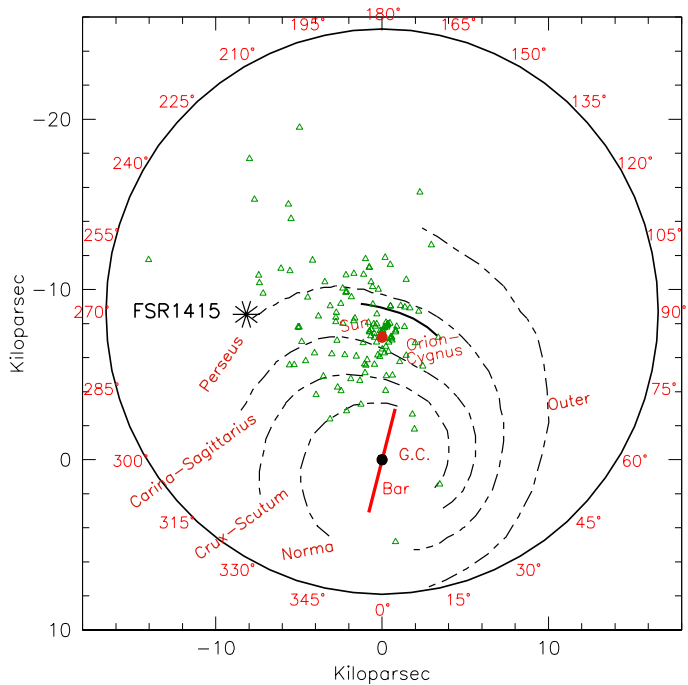}
\caption{A schematic view of the Milky Way as  seen from its North
pole showing the  4 spiral arms, the Galactic   centre, and the  Sun's
position (with R$_{\odot}=7.2\pm0.3$\,kpc).   Also plotted  are a WEBDA
open cluster compilation (limited  to an age of  $\ge 1$ Gyr)  and the
position of FSR\,1415 as derived in this paper.}
\label{f_XY}
\end{figure}

To place FSR\,1415 in    a bigger framework, in   Figure~\ref{f_XY} we
display  a  schematic   view  of the  Milky    Way  Galaxy  (from
\citealt{drimmel06}, see also \citealt{vall08}) in which its  major 
features, spiral arms and bar, are outlined.
Superimposed (as  open triangles) we show the  positions of  125 known
old ($\ge1$ Gyr)    open  clusters, taken  from  the   WEBDA  database
(\citealt{Merm1996}\footnote{ The Galactocentric  coordinates  of
the   WEBDA open    cluster   sample  have been calculated    assuming
d$_{\odot}=7.2$ Kpc as derived in Bica et al. (2006).} ).
Their spatial distribution shows a  clear  detection ``bias'', in  the
sense of a concentration towards the Sun and the Solar circle. In this
context, FSR\,1415 stands out as a  distant cluster in quadrant III of
the  Galaxy, located  at  R$_{\rm GC}=11.81$   kpc  from the  Galactic
centre\footnote{With  R$_{\odot}=8.0$ instead  of R$_{\odot}=7.2$  the
resultant distance  from  the Galactic centre  would  be $12.18\pm~0.34$
kpc}.
Figure~\ref{f_XY} shows that there are  a handful of old open clusters
in this  direction, and that more  old clusters await to be identified
in quadrant IV.  Besides their age, another  interesting aspect of old
clusters  such  as  FSR\,1415 is  their  identification  as outer disk
clusters.  Indeed, the  determination  of  their metallicity  will  be
crucial in extending analysis of reported metallicity gradients far to
the Milky Way outskirts.

%
%

\subsection{Structure}
\label{Struc}

Structural  parameters  of FSR\,1415  were  derived  from the  cluster
stellar radial density profile  (RDP), which consists of the projected
number-density  of  stars  around  the  cluster centre.   The  RDP  of
FSR\,1415  was built  by merging  the observed  VLT photometry  in the
inner  region,   and  the   2MASS  data  in   the  outer   part  (e.g.
\citealt{BB07}). The field background was obtained from the 2MASS data
and  subtracted.   The background-subtracted  2MASS  profile has  been
scaled  to  match  the  VLT  one  by  multiplying  the  2MASS  RDP  by
$28.9$. The VLT  profiles of FSR\,1415 are limited  to $K=19.0$, which
corresponds to  the stellar mass $\approx0.93$  M$_{\odot}$. We employ
the observed photometry because the  artificial stars do not take into
account dynamical  effects. The VLT/MAD  central surface density  is a
factor $\approx30$  higher than  that of the  2MASS profile,  which is
consistent with the $\approx4$\,mag deeper VLT photometry.

\begin{figure}
\centering
\includegraphics[width=6.7cm]{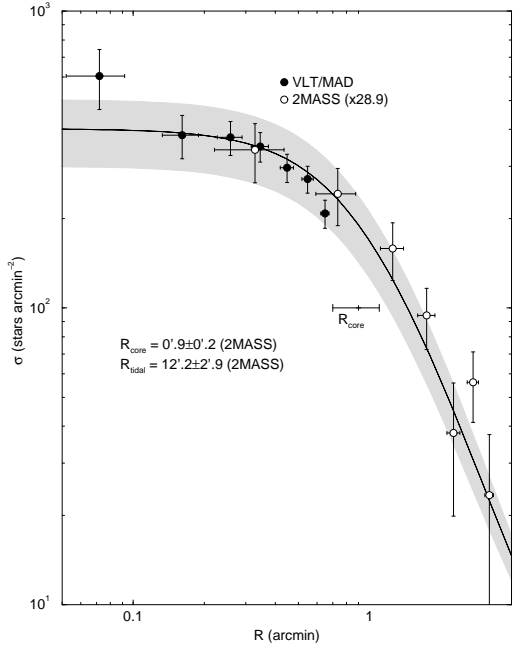}
\caption{VLT/2MASS merged stellar density profile as a function of the
projected distance  to cluster centre.  The observed set of  stars was
employed.  A  3-parameter  King  profile fit  is  superimposed  (solid
line). Shaded area corresponds to the $1\sigma$ fit uncertainties. The
core and tidal radii have been derived from the 2MASS data.}
\label{f_RDP}
\end{figure}

In  Fig.~\ref{f_RDP} we  show  the merged  VLT/2MASS density  profile.
Star clusters  have RDPs  that can be  described by  some well-defined
analytical  profile.   The  most  widely  used  are  the  single-mass,
modified   isothermal  sphere   of   \citet{King1966},  the   modified
isothermal sphere of \citet{wilson1975}, and the power-law with a core
of \citet{elson1987}.
These functions are characterized by different sets of parameters that
describe the cluster structure.  Since we are working with a RDP built
from different photometric sets and that our goal here is basically to
derive reliable structural parameters for FSR1415, we fit the RDP with
the             analytical             3-parameter            function
$\sigma(R)=\sigma_0\left[\frac{1}{\sqrt{1+(R/R_c)^2}}                 -
\frac{1}{\sqrt{1+(R_t/R_c)^2)}}\right]^2$,  where  $\sigma_0$  is  the
central number-density  of stars, and \rc\  and \rt\ are  the core and
tidal radii, respectively.
This  function is similar  to that  introduced by  \citet{King1962} to
describe  the surface  brightness  profiles in  the  central parts  of
globular clusters.   Moreover, as discussed  in \citet{BonattoBica081}
and \citet{BonattoBica082}  RDPs built with  photometry as deep  as in
the present case produce very robust structural parameters when fitted
with the  3-parameter King-like function.   In such cases,  RDPs yield
similar parameters as with surface brightness profiles.

The results  of   the  fit parameters   indicate  a  core  radius   of
$\rc=0\farcm9\pm~0\farcm2$  (corresponding to $2.6\pm~0.6$\,pc), and a
tidal radius of $\rt=12\farcm2\pm~2\farcm9$ (or $35\pm~8$\,pc).  Thus,
most of  the VLT/MAD observations  are restricted to within  the core.
The           resulting          central           density          is
$\rm\sigma=462\pm118\,stars\,arcmin^{-2}$
($\rm55\pm14\,stars\,pc^{-2}$).   With  respect to the 3-parameter
King  fit we  note that  the innermost  RDP point  at $R\sim0\farcm07$
($\sim0.2$~pc) shows  a marginal $1\sigma$ stellar  density excess.
Although  marginal  in  the  case  of FSR\,1415,  such  a  feature  is
characteristic    of     post-core    collapse    globular    clusters
(\citealt{Trager95}), and may indicate an advanced dynamical evolution
state  in the  core  of FSR\,1415  (see Sect.~\ref{LumMasFunc}).  The
presence of post-core  collapse features in RDPs of  open clusters has
been previously  detected, for  instance, in the  $\sim1$\,Gyr cluster
NGC\,3960 (\citealt{BB06}).

\subsection{Luminosity and mass functions}
\label{LumMasFunc}

The observed  luminosity and mass  functions (LF, MF respectively) for
the $J$ and $K$ bands were built using the parameters  in Table~2.  We
applied   a  colour-magnitude  filter   to    the  observed photometry
(Fig.~\ref{f_cmds}) to minimize contamination (e.g. \citealt{BB07}).

\begin{figure}
\centering
\includegraphics[width=8cm,height=8cm]{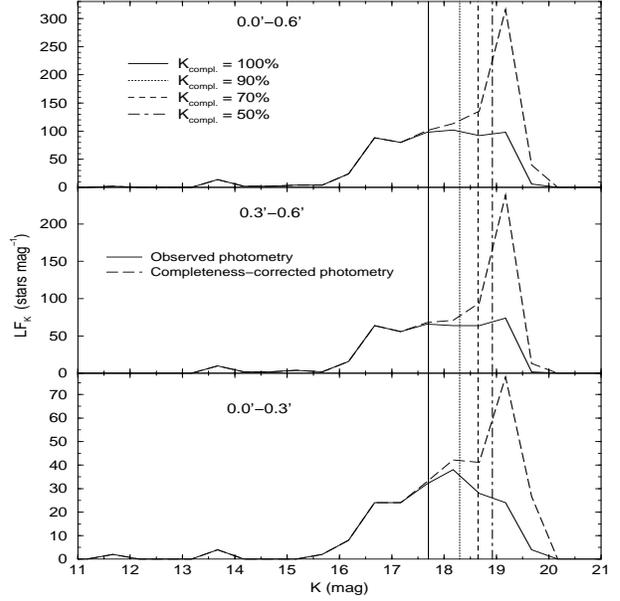}
\caption{Luminosity functions in the $K$ band for an inner 
(bottom  panel), intermediate  (middle  panel), and  wide (top  panel)
regions. These regions  are inside the  core.  Completeness values for
selected magnitudes are shown as vertical lines. LFs with the observed
(heavy-solid  line)   and  completeness-corrected  (heavy-dashed line)
photometry are shown   for  comparison.   }
\label{f_LF}
\end{figure}

Fig.~\ref{f_LF} shows the LFs measured in the $K$ band for the regions
$0\farcm0-0\farcm3$, $0\farcm3-0\farcm6$, and $0\farcm0-0\farcm6$.  It
is interesting  to note that  these LFs correspond to  spatial regions
contained within  the core (Sect.~\ref{Struc}),  and differences among
them  might be  related to  dynamical  effects inside  the core.   The
actual number  of member (i.e.  decontaminated) stars included  in the
computation of the $R<0\farcm6$ LF is  321, 307 of which are MS stars,
while the remaining are evolved.
Granted the isochrone solution derived in Sect.~\ref{age}, we estimate
(for  the observed  magnitude  range) that  the  cluster stellar  mass
amounts to 390\,\ms\ (367\,\ms\ as MS and 23\,\ms\ as evolved stars).
Since completeness  is an important  issue (Sect.~\ref{PhotCompl}) for
fainter  stars, we  show the  completeness levels  for 4  selected $K$
magnitudes.
Besides   the  observed   LFs,   we   also  show    the  corresponding
completeness-corrected ones. The observed LFs do  not appear to change
significantly within    the   core.    The  same     applies  to   the
completeness-corrected LFs.

\begin{figure}
\centering
\includegraphics[width=8.5cm,height=10cm]{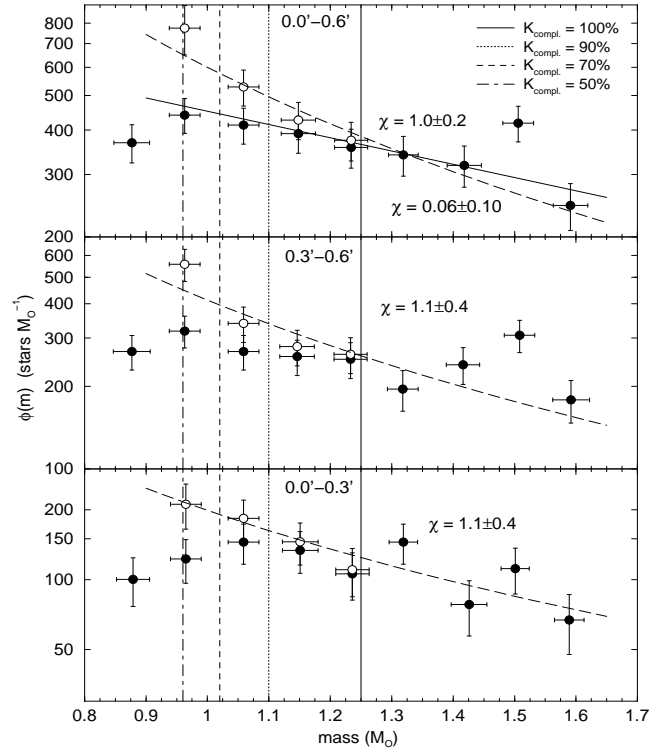}
\caption{Mass functions for the total (top panel), intermediate (middle)
and inner   (bottom) regions.   The   observed (filled   symbols)  and
completeness-corrected MFs (empty) are used. The mass values where the
100\%, 90\%, 70\%,    and  50\% K-completeness  occur   are  indicated
(vertical lines).  Fits  of  $\phi(m)\propto m^{-(1+\chi)}$ are   also
shown.}
\label{f_MF}
\end{figure}

\begin{table}
\centering
\caption[]{Parameters of FSR\,1415, as derived in this paper. We assume
$R_{\odot}=7.2$ kpc.}
\label{t_param}
\renewcommand{\tabcolsep}{.8mm}
\renewcommand{\arraystretch}{1.3}
\begin{tabular}{lr||lr}
\hline
parameters & values & parameters & values\\
\hline
($m-M$)$_{\rm K}$  & 15.20$\pm$0.10    & d$_{\odot}$       & 8.59$\pm$0.45 kpc \\  
($m-M$)$_{\rm J}$  & 15.94$\pm$0.10    & $R_{\rm GC}$   & 11.8$\pm$0.5 kpc     \\   
E${\rm (J-K)}$     & 0.72$\pm$0.02     & $X_{\rm GC}$   & -8.2$\pm$0.1 kpc     \\  
E${\rm (J-H)}$     & 0.46$\pm$0.02     & $Y_{\rm GC}$   & -8.5$\pm$0.5 kpc     \\
E${\rm (B-V)}$     & 1.47$\pm$0.06     & $Z_{\rm GC}$   & -0.3$\pm$0.1 kpc      \\    
A$_{\rm V}$        & 4.56$\pm$0.11     & Mass$_{Obs.}$    & 390 $\ms$ \\    
Age                & 2.5$\pm$0.7 Gyr   & Mass$_{extrap.}$ & $2-3\times10^4\,\ms$ \\   
$[Fe/H]$           & solar             & \rc            & $2.6\pm0.6$ pc    \\
($m-M$)$_{\circ}$  & 14.67$\pm$0.11    &  \rt              &    $35\pm8$ pc  \\

\hline
\end{tabular}
\end{table}
%
%
%
%
%

Mass  functions for  the  same spatial   regions   as the LFs were
constructed  by  combining  the two  $J$   and $K$  independent MFs  
(see, e.g. \citealt{BB05}).
The MFs    were   computed for the   mass    range  $m>0.87\,\ms$,
corresponding  to  $J\la20.2$ and  $K\la19.3$. These are shown in
Fig.~\ref{f_MF}, where the relatively small MF error bars reflect 
the statistical significance of the star counts. At first  sight, the
observed MFs  are flat,   showing  small  differences between the    2
innermost regions.
There   is  however an  important   drop  towards the  low-mass stars.
Although in principle this might be related to mass segregation inside
the core, nevertheless, crowding and completeness should be taken into
account before a similar conclusion can be established.
To this purpose we also built the completeness-corrected MFs, (as done
for the LFs,  Fig.~\ref{f_LF}).  As expected, a significant difference
between the  observed and the  completeness-corrected  MFs is observed
for masses below $1.1\,\ms$.

For a  better description  of the cluster  MFs we  fit these  with the
function $\phi(m)\propto m^{-(1+\chi)}$, carried out for masses higher
than $m>0.96\,\ms$ ($J\simeq19.8$, $K\simeq18.9$).
The lowest-mass  point, at $\approx0.88\,\ms$, is exceedingly affected
by completeness.  The fit to the  observed $0\farcm0-0\farcm6$ MF (top
panel) provides a  slope of $\chi=0.06\pm0.10$,  significantly flatter
than the $\chi  =  1.35$ of  \citet{Salpeter55} initial mass  function
(IMF).     However,   when  the   same   fit     is  applied  to   the
completeness-corrected     MF,  the  resultant     slope  increases to
$\chi=1.0\pm0.2$, thus  steeper  than the  non-corrected MF but  still
flatter than Salpeter's.
For  the innermost  region the  fit to  the  completeness-corrected MF
gives  $\chi=1.1\pm0.4$,  which is  the  same  slope  derived for  the
intermediate region.  Within uncertainties, the completeness-corrected
MF slopes are  uniform throughout the sampled region  within the core,
which suggests that we  are not detecting significant differences
in mass  segregation effects in such  a small spatial  scale,  at
least for stars more massive than $m\approx1\,\ms$.
It may be that  a dynamical effect such  as mass segregation cannot be
detected given  the relatively high stellar mass  range covered by the
present observations, basically $\ga1\,\ms$, and that deeper data (and
lower masses stars) is needed to ascertain this issue.
In any case,  the relatively flat core  MF derived in Fig.~15 suggests
that dynamical effects have affected the core as a whole.

%
%
%

The  total stellar mass in the core can be estimated by using the
MF derived  for the  region $0\farcm0-0\farcm6$, and  extrapolating it
down to the H-burning mass limit, $0.08\,\ms$.
This procedure may over-estimate the  total core stellar mass since it
does not account for the evaporation  of low-mass stars. Nevertheless,
one can provide an upper limit estimate to the cluster mass.
With the observed MFs, we derive  an extrapolated stellar mass of
$730\,\ms$. Since  we are sampling  $\sim2/3$ of the core  radius, the
(observed)  core   mass  would  be   about  $2000\,\ms$.  \citet{BB05}
estimated that for Gyr-class OCs, in general, the extrapolated cluster
mass is  about 10 times  that of the  core. Thus, the  total (observed
 and extrapolated)  stellar  mass of  FSR\,1415  would be  about
$2\times10^4\,\ms$, which is comparable  to those of low-mass globular
clusters (\citealt{Harris96}, and references therein).

When applying the same arguments to the completeness-corrected MF, the
total   (upper  limit)   cluster  mass   is  estimated   to  be
$\sim3\times10^4\,\ms$,  i.e.   $\sim50\%$  higher than  the  observed
value. Once again, such a high mass estimate can be accounted for
by the facts that {\em (i)} the cluster is significantly more extended
than the VLT/MAD-sampled region (Fig.~\ref{f_RDP}), and {\em (ii)} the
MFs  were extrapolated down  to 0.08\,\ms\  stars (while  the observed
mass   range  corresponds   to  stars   more  massive   than  $\approx
0.9\,\ms$).
In any  case, both mass estimates  are consistent with  the large core
($\rc\approx2.6$\,pc) and  tidal ($\rt\approx35$\,pc) radii.  Overall,
the above  mass and  radii estimates indicate  a massive  open cluster
which  shows  evidence  of  an  advanced dynamical  state,  given  its
relatively flat core MF.  Thus, and in spite of being an outer massive
disk cluster,  FSR\,1415 may present features  of dynamical evolution,
suggesting  that it  may underwent  significant  encounters with
GMCs and other tidal effects.

\section{Concluding remarks}
\label{Conclu}

The  present   study  shows  that  the   cluster  candidate  FSR\,1415
(\citealt{FSR07}) is  confirmed as a genuine old  open cluster.  Thus,
the Froebrich et  al.  catalogue appears to be  an important source of
interesting clusters candidates and is worthy of further examination.

The high resolution and deep VLT/MAD $J$ and $K$ data of FSR\,1415 have
allowed an accurate determination of  basic cluster parameters.
It is of   great importance to  obtain similar  high quality data  for
other outer disk open clusters and complement these with spectroscopic
metallicity    determinations.  Consequently,   one  can explore   the
age-metallicity-position dependencies  all   across  the Galaxy.    But
perhaps first of all we need to overcome the current selection bias of
old open clusters in the III and IV Galaxy quadrants.

\section*{Acknowledgments}

The  anonymous referee  is acknowledged  for helpful  suggestions.  We
thank the  ESO Adaptive Optics  group, in particular  Enrico Marchetti
and Paola  Amico.  We acknowledge  partial financial support  from the
Brazilian  agencies  CNPq and  FAPESP.   SO  acknowledges the  Italian
Ministero dell'Universit\`a e  della Ricerca Scientifica e Tecnologica
(MURST).  YM wishes  to thank Marco Gullieuszik and  Simone Zaggia for
useful discussions on the MAD performance.


\centering{{$_{{4~8~15~16~23~42}}$}}

\label{lastpage}
\end{document}